\documentclass[aps,prb,twocolumn,showpacs]{revtex4-1}

\usepackage{graphicx}
\usepackage{dcolumn}
\usepackage{bm}
\usepackage[normalem]{ulem}

\begin{document}

\title{Zero-temperature phase diagram of the second layer of  $^{\bf 4}$He 
adsorbed on graphene}  

\author{M. C. Gordillo}
\affiliation{Departamento de Sistemas F\'{\i}sicos, Qu\'{\i}micos 
y Naturales, Facultad de Ciencias Experimentales, Universidad Pablo de
Olavide, Carretera de Utrera km 1, 41013 Sevilla, Spain}

\author{J. Boronat}
\affiliation{Departament de F\'{\i}sica i Enginyeria Nuclear, 
Universitat Polit\`ecnica de Catalunya, 
Campus Nord B4-B5, 08034 Barcelona, Spain}

\date{\today}

\begin{abstract}
The phase diagram at zero temperature of $^4$He adsorbed on an helium
incommensurate triangular  solid on top of a single graphene sheet has been
obtained using the   diffusion Monte Carlo  method. We have found that, in
accordance with  previous experimental and simulation results for graphite,
the ground state of $^4$He on  this setup is a liquid that, upon
compression, transforms into a triangular solid. To define the stability
limits of both liquid and solid phases, we considered not only the
adsorption energies of the atoms located on the second layer but the
average energy of the atoms in both layers.  Our results show that the
lower density limit for a stable liquid in the second layer is  0.163 $\pm$
0.005 \AA$^{-2}$ and that   the lower limit for the existence of an
incommensurate solid on the second layer is 0.186 $\pm$ 0.003 \AA$^{-2}$.
Both values are in  overall agreement with the results of torsional
oscillator experiments and heat capacity measurements on graphite. The 
4/7 and 7/12 registered solids  are found to be metastable with respect
to triangular incommensurate arrangements of the same density.    
\end{abstract}
 

\maketitle

\section{INTRODUCTION}

Graphene is a novel form of carbon in which the atoms are located in the
vertices of  a two-dimensional honeycomb
lattice.~\cite{science2004,pnas2005} This means that it could be thought as
the result of isolating each of the multiple layers that conforms a
graphite structure, or as the flat counterpart of a carbon nanotube.  In
principle, graphene could be used as a gas adsorbent at very low
temperatures in the same way than graphite  (see for instance Ref.
\onlinecite{colebook}), or a carbon nanotube,~\cite{vilchesscience} but up
to now this  has not been experimentally realized.   The only studies on
that surface are computer
simulations.~\cite{prl2009,prb2010,prb2011a,prb2011b}  All these works
consider only  the first layer of a quantum species ($^4$He, H$_2$)
adsorbed on its surface,  and their results indicate that graphene behaves
as adsorbent basically like graphite.  The main difference is an almost
constant offset of the binding energy per particle due  to the presence of
more carbon atoms in the graphite case. 

In the present work, we are interested in the behavior of $^4$He adsorbed
on top  of a helium triangular solid resting on a single graphene layer,
i.e., in the second layer  of $^4$He on graphene. Since there are not
experimental results on graphene,  the only data to compare to will be  the
heat capacity,~\cite{grey,grey2} third sound,~\cite{chan} and torsional
oscillator~\cite{cro,cro2} measurements on  graphite. And the same applies
to previous Green's function Monte Carlo (GFMC)~\cite{Whitlock} and  path
integral Monte Carlo (PIMC)~\cite{manousakis4,manousakis5,boninx}
calculations on this system: all are performed for $^4$He on graphite. One
of our aims in the present work is to check if  the phase diagrams on
graphite are  similar to those on graphene, as happened to the first layer
of helium.          

The picture that emerges from experimental and simulation results of $^4$He
on graphite indicates  that, when the total density increases, there is a
promotion to a second layer. This transition density was found to be in the
range 0.115 -- 0.12 \AA$^{-2}$ in neutron~\cite{carneiro} and heat capacity
measurements,~\cite{polanco,grey,grey2} in good agreement with
PIMC~\cite{boninx} (0.1140 $\pm$ 0.0003 \AA$^{-2}$), and    
GFMC~\cite{Whitlock} (between 0.115 and 0.118 \AA$^{-2}$) calculations. 
From there up, according to heat capacity data,~\cite{grey2} there is a
stable liquid phase in  the range 0.16 -- 0.19 \AA$^{-2}$; this interval
includes the one inferred from torsional oscillator
experiments~\cite{cro,cro2} (0.174 -- 0.19 \AA$^{-2}$).   From  0.19
\AA$^{-2}$ up, an incommensurate solid is found, whose stability  limit
ends with the helium promotion to the third layer at densities ranging from
0.204 \AA$^{-2}$ (Refs.\onlinecite{chan,cro,cro2}) to 0.212 \AA$^{-2}$
(Ref. \onlinecite{grey2}). Experimental~\cite{grey2} and 
theoretical~\cite{manousakis4,manousakis5} data also suggest the existence
of a commensurate  4/7 phase (registered with the first layer
incommensurate solid) in this second layer,  in an analogy to what happens
to $^3$He on the same substrate,~\cite{fukuyama} even though other 
calculations~\cite{boninx} contradict these findings. The existence of a 
7/12 registered solid was also studied in Ref. \onlinecite{boninx} and
found unstable.       
 
All  previous simulations that went beyond considering the second layer
of $^4$He  on  graphite as a purely two-dimensional system were PIMC 
calculations.~\cite{manousakis4,manousakis5,boninx}
The study of the phase boundaries with PIMC is difficult 
since to check the relative  stability of the different atom arrangements
free energy calculations~\cite{he4} have to be done and the primary output
of finite temperature calculations is the energy, not the free
energy.   On the other hand, the diffusion Monte Carlo (DMC) technique
appears to be ideally suited for that purpose.  First, it is 
a ground-state method which works at $T=0$.
And second, it allows us to introduce at will the  particular phase or set
of phases we are interested in through an appropriate importance sampling.
One arrangement is preferred over another with the same density if the
energy per particle in the former  is smaller than in the latter. This
simple prescription, together with the  use of double tangent Maxwell
constructions,  permit us to establish with precision the stability
boundaries of  the different phases proposed. 

In this work, we report
the results of performing different series of DMC  calculations to obtain
the ground-state phase diagram of the second  layer of $^4$He adsorbed on graphene. The
next section is devoted to the description of the technique and 
approximations used to do so. After that, Section III  shows the results
obtained, leaving the conclusions for Section IV.      

\section{METHOD}

Our microscopic study of the second layer of $^4$He on graphene is based on
the DMC method. Nowadays, DMC is a standard tool that allows for an exact
calculation of ground-state properties of boson systems, within some
statistical uncertainties, by solving stochastically the multi-particle Schr\"odinger
equation.~\cite{boro94} In order to reduce the variance to a manageable
level, one introduces the usual importance sampling strategy through a guided
diffusion process. The drift force is intended for focusing the sampling to
regions where one reasonably expects that the wave function of the system is
large. Technically, this is implemented by introducing a guiding wave function
that avoids interparticle distances smaller than the core of the
interactions, localizes particles close to the adsorbing surface, and fixes
the phase (liquid or solid) of the system. The model we have used in the
present study is
\begin{eqnarray}
\Phi({\bf r}_1,\ldots,{\bf r}_N) & = & \Phi_J({\bf r}_1,\ldots,{\bf r}_N)
\Phi_1({\bf r}_1,\ldots,{\bf r}_{N_1})  \nonumber \\
& & \times \ \Phi_2({\bf r}_{N_1+1},\ldots,{\bf r}_N) \ ,
\label{phitot}
\end{eqnarray}
where  $N$ is the total number of atoms and $N_1$ the fraction of them
located in the first layer. The first term is a Jastrow factor that
accounts for the $^4$He-$^4$He dynamical correlations induced by their
interparticle interaction,
\begin{equation}
\Phi_J({\bf r}_1,\ldots,{\bf r}_N) = \prod_{1=i<j}^{N} \exp \left[-\frac{1}{2} 
\left(\frac{b}{r_{ij}} \right)^5 \right] \ .
\label{sverlet}
\end{equation}
In Eq. (\ref{sverlet}), $b$ is a variational parameter that has been
already optimized in previous
simulations of $^4$He in various setups~\cite{boro94,prl2009,giorgini} and
found to be $3.07$ \AA. The second and third terms in the r.h.s. of Eq.
(\ref{phitot}) are introduced to describe approximately the first and
second layer, respectively. Explicitly,
\begin{eqnarray}
\lefteqn{\Phi_1({\bf r}_1,\ldots,{\bf r}_{N_1})  =  \prod_{i=1}^{N_1}
\Psi_1(z_i)} \nonumber \\ 
& & \times \prod_{i,I=1}^{N_1} \exp \{-a_1 [(x_i-x_I^{(1)})^2 +
(y_i-y_I^{(1)})^2] \} \ , \\
\lefteqn{\Phi_2({\bf r}_{N_1+1},\ldots,{\bf r}_N)   =  \prod_{i=N_1+1}^{N}
\Psi_2(z_i)} \nonumber \\
&  & \times \prod_{i,I=N_1+1}^{N} \exp \{-a_2 [(x_i-x_I^{(2)})^2 + (y_i-y_I^{(2)})^2] \} 
\ .
\label{phicapa}
\end{eqnarray} 
In $\Phi_1$, the Gaussian terms define the triangular solid that comprises the first
layer, where the coordinate set $(x_I^{(1)},y_I^{(1)})$ are the lattice
sites; these positions can be varied to consider  different densities. 
The optimal value for the parameter $a_1$, that defines the strength of the
localization factor, has been taken from DMC simulations of the first 
layer.~\cite{prl2009} As in that work, the parameter $a_1$ is varied  
linearly with density between $a_1=0.30$ \AA$^{-2}$ at 0.08 \AA$^{-2}$ and
$a_1=0.77$ \AA$^{-2}$ for 0.128 \AA$^{-2}$. On the other hand, $\Psi_1(z)$ is the solution of the
one-dimensional Schr\"odinger equation for a single atom
moving  perpendicularly to the graphene plane with a potential that is
the averaged out version in the $z$ axis  of all the Lennard Jones
interactions between all the carbon atoms and the $^4$He atom. This is
exactly what was  made in Refs. \onlinecite{prl2009} and
\onlinecite{Whitlock}.   

$\Phi_2$ in Eq. (\ref{phicapa}) accounts for the properties of the second
layer. The function $\Psi_2(z)$ is the solution
of the Schr\"odinger  equation for a single atom on top of both graphene and
a fixed helium layer of  density 0.11 \AA$^{-2}$. It has a maximum at
a distance of $5.59$ \AA \ from the position of the  graphene layer, and it
has been 
used for all the calculations involving the different phases on the second
layer. As in the first layer, the different solid phases in the second
layer  have been modeled by
defining their corresponding  lattice positions $(x_I^{(2)},y_I^{(2)})$ and by their $a_2$
parameters, obtained variationally for each considered density. This
optimization leads to the same density dependence that the one obtained for
the first layer.  If the upper layer is a liquid, $a_2 = 0$. 

The helium-helium interactions in both
layers were modeled by an  Aziz potential,~\cite{aziz} and the
carbon-helium ones by a Lennard-Jones interaction.~\cite{cole} All the carbon atoms in
the graphene layer have been considered individually, i.e., full corrugation
effects have been taken into account. We also tested an alternative form of the carbon-helium
interaction that is explicitly anisotropic,~\cite{carlosandcole} and the results were
found to be very similar to the ones obtained with the isotropic
Lennard-Jones potential. The only difference was a decrease in the energy
per particle in the range 0.5 -- 1 K, depending on the density of the
particular arrangement considered. The stability limits of the
phases remained independent of this choice for the pair carbon-helium
potential as expected because the second layer is much less influenced by
corrugation effects than the first one.   

When the second layer phase is  liquid or a registered structure, a
rectangular simulation cell, defined by the triangular solid that comprised
the underlying helium sheet, is used. However, if the upper structure is an
incommensurate arrangement, there is in principle a mismatch between the
first and  second crystallographic structures.  To solve the problems
created by this situation, we define a given rectangular simulation cell
for the first layer,  and located on top of it the largest piece of the
incommensurate solid that fitted within those limits. To avoid the effect
of second layer empty fringes in the global interactions,  the periodic
images of  the second layer are taken into account to calculate the total
energy. Working in this way, we can consider any upper or lower densities,
not needing to confine ourselves to quasi-commensurate arrangements.~\cite{manousakis5}  
In all calculations, all the positions of the
particles, both in the first and second layer, were allowed to move during
the simulations, i.e, zero-point motion was considered for all helium
atoms.

\section{RESULTS}

The first concern about the  second layer of
a given adsorbate is  the total density at which that layer starts to form.
One way to answer this question is to determine the chemical potential of a
single atom adsorbed on top of a helium layer and compare it to the same
parameter for the atoms that constitute   the lower solid helium layer. The
promotion will take place at the particular first layer density at which
both   chemical potentials are equal. This is the procedure used previously
in graphite,~\cite{Whitlock,boninx}   giving  
a first layer density between 0.114 and 0.118 \AA$^{-2}$. 
Our approach relies on the same theoretical grounds but it is slightly
different from a technical point of view: to determine the
lowest second layer density for which a liquid phase is stable, we 
perform a Maxwell construction between  the simulation results for the
energy per particle in a one-layer setup and the ones for an arrangement
with a liquid on top of a quasi-two-dimensional solid. 

\begin{figure}[b]
\begin{center}
\includegraphics[width=7.5cm]{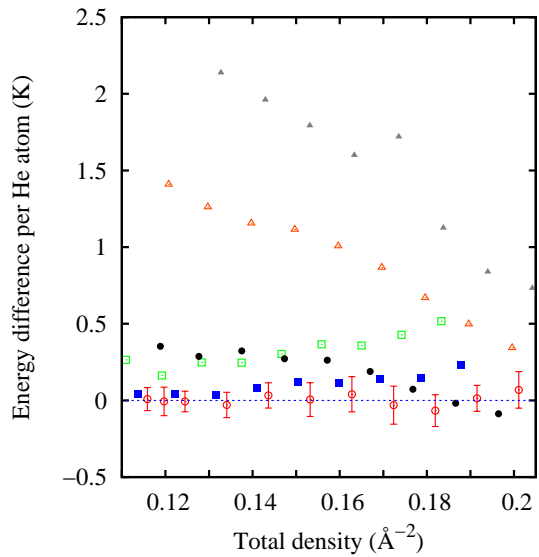}
\caption{(Color online) Relative energy per particle values with respect to
a fourth-order polynomial fit to the energy  per particle of an structure in
which the underlying solid has a density of 0.115 \AA$^{-2}$ (dashed line).
From top to bottom, we show liquid layers on top of solids of 0.1225
\AA$^{-2}$ (full triangles), 0.12 \AA$^{-2}$ (open triangles), 0.1175
\AA$^{-2}$ (full circles), 0.115 \AA$^{-2}$ (real simulation results minus
the values obtained form the fit at the same densities, open circles),
0.1125 \AA$^{-2}$ (full squares) and 0.11 \AA$^{-2}$ (open squares).     }
\label{energy1}
\end{center}
\end{figure}

The first step in the present case is to determine which is the density of the 
underlying solid for which the
energy per atom (including both layers) is the lowest one. 
This can be done with the help of Fig.~\ref{energy1}.
There, we plot the relative energy
per particle for different first layer solids as a function of the total density.
We have used relative values instead of absolute ones due to the small  energy
differences between the possible configurations. We choose as reference
values for each density  the ones    for an upper liquid layer in
which the first solid helium layer has a fixed density of 0.115 \AA$^{-2}$. What
we present in   Fig.~\ref{energy1} is the difference between the energy
per atom for a given structure and a fourth-order polynomial fit to the
values corresponding to solid density 0.115 \AA$^{-2}$. This means that if the energy
per particle for a particular arrangement is lower than the one  for a
liquid on top of a solid of 0.115 \AA$^{-2}$ with the same total density,
it will be represented by a negative value in  Fig.~\ref{energy1} and
vice versa. The real simulation results for the reference setup are displayed
as open circles with their error bars, to be compared with the dashed line
that indicates the fit to them. The error bars for other structures and
densities  are similar to those and not displayed for clarity. What we
can see is that the relative energies   for all other setups, with
underlying solid densities ranging from 0.11 (open squares) to 0.1225
\AA$^{-2}$ (full triangles), are always greater than zero, i.e., the
system with the smallest energy is a liquid on top of a triangular
solid of density 0.115 \AA$^{-2}$.   The only possible exception would be a
liquid with $\sigma >  0.0725$ \AA$^{-2}$ on top of a solid of density 0.1175
\AA$^{-2}$ (a total density $\sim$ 0.19 \AA$^{-2}$). However, at these
densities the liquid phase on the second layer is no longer stable (see
discussion below).  We can see also that in the range 0.11 -- 0.1175 \AA$^{-2}$ the
energy differences for the full structures are very small, even lower than
1 K, pointing to the difficulty of a precise determination of the promotion
density.

\begin{figure}[b]
\begin{center}
\includegraphics[width=7.5cm]{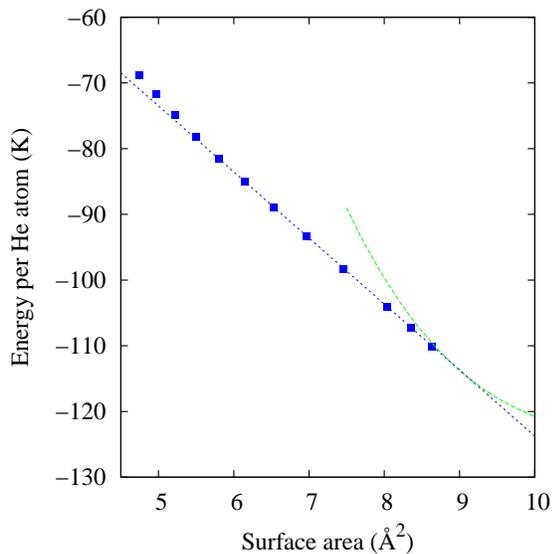}
\caption{(Color online) Energy per particle for a structure with a
single layer triangular solid  (dashed line) and a second layer liquid on
top of a solid of density 0.115 \AA$^{-2}$ (full squares) versus the
inverse of the total helium density. The common tangent of a Maxwell
construction is represented by a dotted line.}
\label{energy2}
\end{center}
\end{figure}

Once we have established from first principles which first layer solid to use, one
should perform a double-tangent Maxwell construction to determine   the
lowest stability limit for a second layer liquid after helium promotion.
The necessary data are displayed in  Fig.~\ref{energy2}. There, we show a
third-order polynomial fit to the energy per particle of single layer
solids of different densities, taken from Ref. \onlinecite{prl2009} (dashed
line), together with the present simulation results (full squares) corresponding to
liquids on top of a first layer of density 0.115 \AA$^{-2}$.  The dotted
line defines the common tangent for the equilibrium configurations in the one
and two layer structures. This means that  the first layer solid is stable up
to the surface area at which the energy per particle curve has the same
slope than the dotted line, and the second layer liquid starts at the
last density at which its curve has also that slope. The intermediate points
correspond to  inhomogeneous second layer liquid arrangements whose
energies per particle are the density weighted averages of the energies of
the structures in equilibrium at the transition. From Fig.~\ref{energy2},
we can state that      the approximate density of the one layer solid in
equilibrium with the liquid is $0.113 \pm 0.001$ \AA$^{-2}$, corresponding
to a
surface  area of $\sim$ 8.8 -- 8.9 \AA$^{2}$, and that the pressure (minus
the slope of the Maxwell construction line) at which the first order 
transition takes place is $\sim$ 10.06 K/\AA$^2$. That density corresponds
to the helium promotion to the second layer and  is compatible with those
of  previous simulations~\cite{Whitlock,boninx} and with neutron
diffraction experiments~\cite{carneiro} on graphite. However, this value is
somehow smaller than the  one inferred from some heat capacity 
measurements~\cite{grey,grey2} (0.12 \AA$^{-2}$) on the same substrate, even though is
compatible with others.~\cite{polanco} Our result points to the prediction
that  the second
layer promotion of $^4$He on graphene takes place at the same density than
on graphite.  

\begin{figure}[b]
\begin{center}
\includegraphics[width=7.5cm]{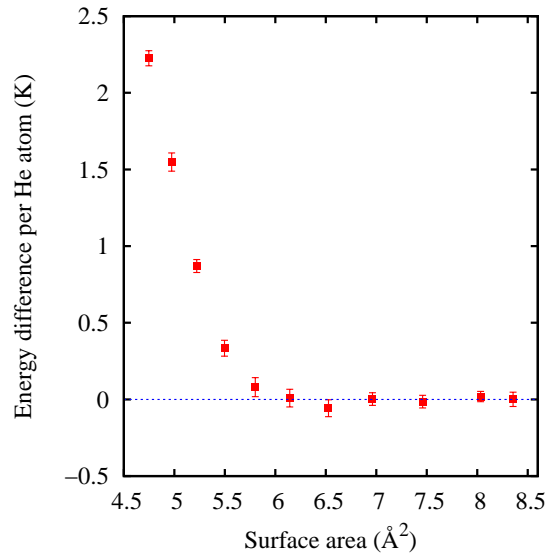}
\caption{(Color online) Energies per particle from the
simulation results displayed in Fig.~\ref{energy2} (full squares) minus
the  corresponding values obtained from the double tangent line, here
represented as a dashed line.}
\label{energy3}
\end{center}
\end{figure}

The lower density limit for the second layer liquid is then the last point
for which the simulation results share the common tangent. This is
difficult to see in Fig.~\ref{energy2}, and for this reason we have displayed
the same data in a slightly different form in Fig~\ref{energy3}. There, we
show the difference between the energy per particle obtained from the
simulations and the  one obtained from the Maxwell construction
line. Obviously, that difference is zero if the points are on top of that
line: the smaller surface area that fulfills this
prescription is  6.14 \AA$^2$, that   corresponds to a density
of 0.163 $\pm$ 0.005 \AA$^{-2}$. The error bar is estimated as half of the
distance between  the simulated points. This value compares favorably with
the experimental data of Greywall~\cite{grey,grey2} ($\sim$ 0.16
\AA$^{-2}$),  but is slightly smaller than the one obtained from torsional
oscillator experiments by Crowell and Reppy in graphite ($\sim$ 0.17
\AA$^{-2}$).~\cite{cro,cro2}  If we subtract the density of the first layer,
the remaining density, $0.048  \pm 0.005$ \AA$^{-2}$ is also compatible
with the strictly two-dimensional calculations of Refs.
\onlinecite{giorgini},\onlinecite{he4} ($\sim$ 0.043  \AA$^{-2}$ in
both cases), and with  PIMC results on graphite,~\cite{boninx}
in which a similar value is given (0.046 \AA$^{-2}$).  In the latter case,
the full three-dimensional structure was used,  with an underlying solid in which the
atomic zero-point motion was taken into account.
However, in Ref.\onlinecite{boninx}  no prediction about the density of the first
layer is given, stating only that the  location of the liquid equilibrium
density was independent of that value. This is probably due to the small
energy differences between  structures with different underlying solid
densities, as commented above.       

\begin{figure}[b]
\begin{center}
\includegraphics[width=7.5cm]{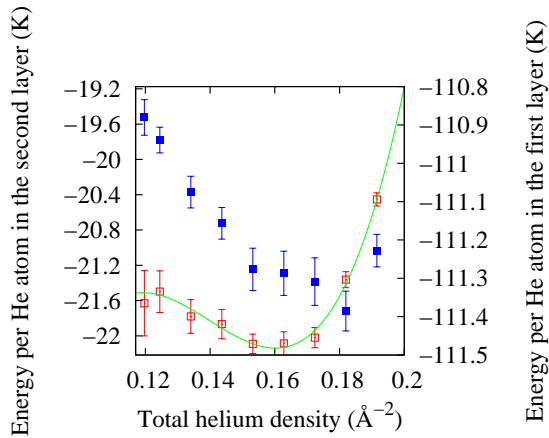}
\caption{(Color online) Energy per particle of $^4$He in the first
(full squares, right scale) and second (open squares, left scale) layer 
for a liquid phase on top of a solid of density 0.115 \AA$^{-2}$.  
}
\label{dobleenergy}
\end{center}
\end{figure}

\begin{figure}[b]
\begin{center}
\includegraphics[width=7.5cm]{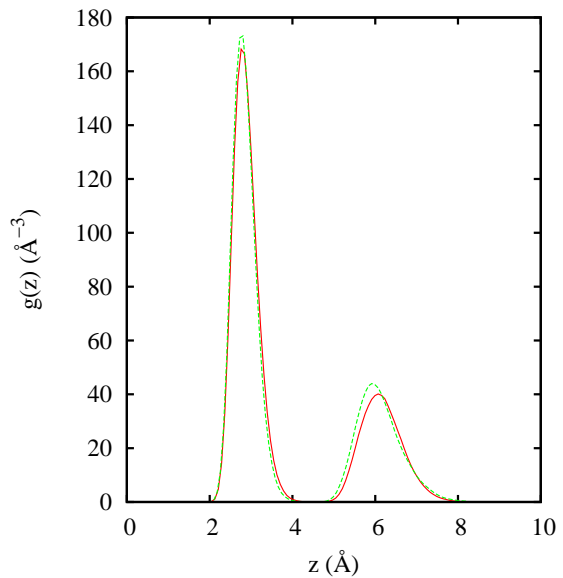}
\caption{(Color online) Density profiles as a function of the
distance perpendicular to the graphene layer for a  second layer liquid 
with total density 0.163 \AA$^{-2}$ (solid line) and a double solid of total
density of 0.1875 \AA$^{-2}$ (dashed line). In the latter case, the
density of the underlying solid is 0.1175 \AA$^{-2}$.}
\label{perfil}
\end{center}
\end{figure}

If we consider only the energy per particle of the helium atoms in the
second layer, we have the isotherm displayed in Fig.~\ref{dobleenergy}. 
There, we show the simulation results (open squares, left scale) together
with a third-order polynomial fit that allows us to estimate that the minimum
energy per particle is $-22.14 \pm 0.04$ K for an equilibrium density of
$0.045 \pm 0.001$ \AA$^{-2}$. This density  is a  bit lower  but compatible
with the result quoted in the previous paragraph ($0.048  \pm 0.005$
\AA$^{-2}$). At least,
part of the difference could be ascribed to  the behavior of the $^4$He
atoms  
closest to the graphene substrate: far from being constant, the energy per
particle decreases up to 0.8 K per particle due to the presence of the
upper liquid layer. This means that to describe this system
probably one has to take into account  both layers. The infinite-dilution
limit  for the second layer at this underlying density is 
$-21.53 \pm 0.01$ K. Previous simulation
results on graphite report values of $-29.6 \pm 0.3$ K (Ref.
\onlinecite{boninx}), and $-29.8$ K for slightly smaller underlying helium
densities. This implies an increase in the binding energy of  a single atom on 
the top layer from graphene to graphite of about 40 \% , versus a
similar 10 \% growing for a single atom on bare graphene or 
graphite.~\cite{prl2009} On the other hand, the energy difference between the
infinite-dilution limit and the energy minimum is $-0.61 \pm 0.04$ K, to be
compared to  $\sim  -0.9$ K of the pure two-dimensional calculation with the
same potential.~\cite{giorgini} This means that the second
layer liquid is not fully two dimensional, as can be seen also with the
help of  Fig~\ref{perfil}. There, we show the density profiles in a
direction perpendicular to the graphene layer of both layers in two
arrangements: the lowest liquid density  limit (full line), and an upper
layer solid of 0.07 \AA$^{-2}$ on top of a triangular solid of 0.1175
\AA$^{-2}$ (dashed line). Far from being delta functions, both layers
display a sizeable width, that is obviously larger for the top one. 

\begin{figure}[b]
\begin{center}
\includegraphics[width=7.5cm]{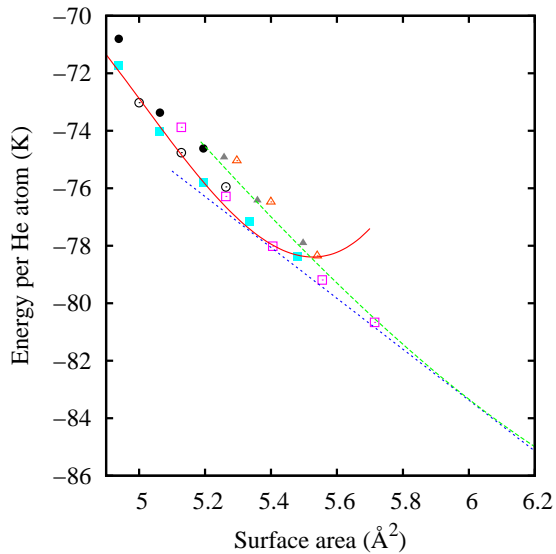}
\caption{(Color online) Energies per particle for different
incommensurate solids versus the inverse of the total density.  The first
layer densities are 0.115 \AA$^{-2}$ (open squares); 0.1175 \AA$^{-2}$
(full squares); 0.12 \AA$^{-2}$ (open circles) and  0.1225 \AA$^{-2}$ (full
circles). The dashed line represents a fourth-order polynomial fit to a
second layer liquid, with underlying density of 0.115 \AA$^{-2}$, while the
dotted line is the double-tangent Maxwell construction. Triangles stand for
the energy per particle for  two commensurate unstable arrangements on top
of different first layer incommensurate solids: full triangles, 7/12
structure; open triangles,  4/7 solid.}
\label{solid}
\end{center}
\end{figure}

If more helium is added, the system evolves to form an
incommensurate second layer solid. The lower density limit of the layer
solid at the transition can be obtained with the help of Fig.~\ref{solid}. 
In this figure, we
display as symbols the energies per particle for different solid
arrangements as a function of the surface area. The error bars are of the
size of the symbols and  not shown for simplicity. The lower value in the
$x$ axis corresponds to  the smaller experimental density at which a
promotion to the third layer has been reported.~\cite{cro,cro2}  The dashed
line is the fourth-order polynomial fit to the simulation data for a liquid
phase, already used in Fig.~\ref{energy1}. The Maxwell construction
between this structure and a second layer solid on top of a triangular 
solid with a density of 0.1175 \AA$^{-2}$ (full squares) is represented by
the dotted line, while the full curve is a fit to those data. 
From Fig.~\ref{solid} it is clear that the upper equilibrium density for
a second layer liquid is around $0.170 \pm 0.005$ \AA$^{-2}$, derived from
a surface area of $\sim$ 5.9 \AA$^2$. We can also see that the open
squares,
representing an incommensurate top solid on a first layer arrangement of
density 0.115 \AA$^{-2}$, are on top of the common tangent line. The upper
surface limit, according to the Maxwell construction, is comprised between 
densities 0.185 and 0.1875 \AA$^{-2}$.  The first value corresponds to a
incommensurate solid of 0.07 \AA$^{-2}$ on top of a 0.115 \AA$^{-2}$,
and the second, to an upper arrangement of the same density but on top of a
first layer solid of 0.1175 \AA$^{-2}$. Unfortunately, the simulation
results do not allow us to discriminate between these two possibilities.
The upper layer density is slightly smaller than the result obtained in previous
PIMC calculations~\cite{boninx} ($0.076 \pm 0.002$ \AA$^{-2}$) and to
a zero-temperature DMC estimation of the melting density in a strictly
two-dimensional crystal ($0.075 \pm 0.001$ \AA$^{-2}$).~\cite{claudi} 
From the difference in the total  density, we can
estimate than the error  bar for the upper density limit is 0.0025
\AA$^{-2}$, i.e., our estimation for the total density is $0.186 \pm 0.003$
\AA$^{-2}$.  This density is  in a reasonable agreement with the value of
0.19 \AA$^{-2}$ that heat capacity data~\cite{grey,grey2} shows for the
appearance of a {\em commensurate} solid, and with the experimental value
of $\sim$ 0.191 \AA$^{-2}$ for the disappearance of     the superfluid
signal in torsional oscillator experiments,~\cite{cro} both experiments on
graphite. If we keep on adding helium,       for total densities larger
than $\sim 0.1950 \pm 0.0025$ \AA$^{-2}$, corresponding to a surface area
of 5.13 \AA$^2$, we can see that the results for  an incommensurate solid
with a first layer density of 0.12 \AA$^{-2}$ (open circles) are on top of
the fit corresponding to a  0.1175 \AA$^{-2}$ solid. Therefore, 
for this density on a different stable two layer solid
phase could be possible, even though  one cannot discriminate between both from energetic
arguments. However, we are sure that  no further compression of
the first layer is possible, since the energy per helium atom  for a double
solid with a lower density of 0.1225 \AA$^{-2}$ is bigger than for its 0.12
\AA$^{-2}$ counterpart (full circles in Fig.~\ref{solid}.  Obviously,
none of these changes can be described by a two-dimensional model since
the density of the underlying solid is different for the possible phases 
phases in equilibrium.

To complete the study of the second layer phase diagram, we have analyzed
the possible  existence of commensurate  structures on
top of the first layer triangular $^4$He solids. The existence of a
$\sqrt{7} \times \sqrt{7}$ phase, equivalent to the 4/7 commensurate arrangement found
experimentally~\cite{fukuyama} in $^3$He  has been 
theoretically~\cite{manousakis4,manousakis5} and experimentally~\cite{grey2} proposed for
graphite. However, a previous calculation of helium on graphite~\cite{boninx}
indicated that the theoretical stability of  this phase was motivated by
the consideration of a completely frozen first layer solid.
 This conclusion is  supported by our data displayed in
Fig.~\ref{solid} as open triangles. The three points shown in the figure correspond to 4/7
structures on top  of first layers of densities 0.115, 0.1175 and 0.12
\AA$^{-2}$; all of them are unstable with respect of a second layer
liquid  of the same total density. The same can be said of the possibility
of a 7/12 phase. The three full triangles in  Fig.~\ref{solid} represent
the energies for that lattice on top of the same first layer solids
than in the previous case. In all cases, the registered solids are also unstable with
respect to the second layer liquid.       

\section{CONCLUSIONS}

Our microscopic results show that $^4$He promotion to a second layer
on graphene starts at a first layer density of $0.113 \pm 0.001$ \AA$^{-2}$. This
phase is in equilibrium with a second layer liquid of density 0.048
\AA$^{-2}$ on top of an underlying solid of 0.115 \AA$^{-2}$. Upon addition
of  more helium, the second layer liquid undergoes a first order phase transition to an
incommensurate second layer solid of density 0.070 \AA$^{-2}$ on top of
either a 0.115 or a  0.1175 \AA$^{-1}$ layer (both options are equally
possible within the accuracy of our calculations). Therefore, the stability
window for a second layer homogeneous liquid is 0.163 -- 0.170 \AA$^{-2}$, 
and that from this last density up, there should be a coexistence zone with
a two-solid structure whose maximum total density is 0.1875 \AA$^{-2}$.
This means  that from 0.170 \AA$^{-2}$ on, the upper layer should be formed
by patches of superfluid liquid together with incommensurate solid
domains.  Those liquid patches would be responsible for the torsional
oscillator signal seen up to 0.191 \AA$^{-2}$ in graphite. No indication of
the width of the transition region can be deduced from the heat capacity
measurements on the same substrate. We are only aware of a value 
0.0031 \AA$^{-2}$ reported in PIMC calculations~\cite{he4} at $T = 0.25$ K, 
five times smaller than the difference estimated here, between 0.185 and
0.170  \AA$^{-2}$. Even though, this is at best a crude approximation, it
is clear that experimentally a transition region should exist, and it
should start at densities lower than the last for which a superfluid signal
can be obtained. Our results also suggest that the  peak that appears in
the heat capacity data of Ref. \onlinecite{grey2} at 0.19 \AA$^{-2}$ could
be ascribed to a transition to an incommensurate  solid, not to a
commensurate one.  

Our simulation results suggest also an explanation for the displacement of
the transition peak from 0.19 to 0.197 \AA$^{-2}$ in heat capacity data: it
could be the result of a further compression of the first layer solid. However,
in our data, that compression is smaller than  the measured in Refs.
\onlinecite{grey,grey2}: a theoretical maximum density of 0.12 \AA$^{-2}$ instead of 
the 0.127 \AA$^{-2}$ density
of the experimental data, a 6\% of difference.  
The experimental difference between the total density at the second layer
promotion (0.12 \AA$^{-2}$) and that of the underlying solid near second
layer compression (0.127 \AA$^{-2}$) is also of 6\% .
The equivalent values obtained from our simulations are 0.113 and 0.12
\AA$^{-2}$, also a 6\% of change.

\acknowledgments
We acknowledge partial financial support from the
Junta de Andaluc\'{\i}a group PAI-205 and Grant FQM-5985, DGI (Spain)
Grants FIS2010-18356 and FIS2011-25275, and Generalitat de Catalunya 
Grant 2009SGR-1003.

\end{document}